\documentclass[english,12pt]{article}
\usepackage[cp1251]{inputenc}
\usepackage{babel}
\usepackage{amsfonts, amsmath,amssymb,graphics}
\newcommand{\be}{\begin{equation}} \newcommand{\ee}{\end{equation}}

\begin{document}
\begin{center}
{\bf The Black Hole Quantum  Entropy  and Its  Minimal Value}\\
\vspace{5mm}Ya.V. Dydyshka \footnote{E-mail:
    yahor@hep.by} and  A.E.Shalyt-Margolin
\footnote{E-mail:
    a.shalyt@mail.ru; alexm@hep.by}\\ \vspace{5mm} \textit{National
Centre of Particles and High Energy Physics, Bogdanovich Str. 153,
Minsk 220040, Belarus}
\end{center}
\rm\normalsize \vspace{0.5cm}

\begin{abstract}
In the paper it is demonstrated that the Schwarzschild black-hole
quantum entropy computed within the scope of the Generalized
Uncertainty Principle (GUP) has a nonzero minimum under the
assumption that for a radius of the black hole the lower limit is
placed, whose value is twice the minimal length.  Such a limit is
quite natural when using, as a proper deformation parameter in a
quantum theory with a minimal length, the dimensionless small
parameter introduced previously by one of the authors in
co-authorship with his colleagues and caused by modification of
the density matrix at Planck’s scales. The results obtained have
been compared to the results of other authors and analyzed from
the viewpoint of their compatibility with the well-known facts and
the holographic principle in particular.
\end{abstract}

\section{Introduction}
The black hole entropy is essential for a quantum field theory in
the curved space-time, for cosmology, physics of the early
Universe, and the like since the appearance of the renowned works
by Bekenstein and Hawking in the seventies of the last century
\cite{Bek1},\cite{Hawk1}. Bekenstein and Hawking have introduced
the black hole entropy in a semiclassical approximation only, i.e.
when material fields are quantized on the basis of the classical
space-time.  Then the following familiar formula for the black
hole entropy is the case:
\begin{equation}\label{GUP4}
S^{BH}=\frac{A}{4l^{2}_{p}}.
\end{equation}
But it is obvious that a change to higher energies (with regard to
the quantum-gravitational effects) necessitates addition of
quantum corrections in the right-hand side of (\ref {GUP4}).   The
importance of this problem has been greatly increased with the
introduction of the Generalized Uncertainty Principle (GUP)
\begin{equation}\label{GUP1}
\triangle x\geq\frac{\hbar}{\triangle p}+ \ell^2 \frac{\triangle
p}{\hbar},
\end{equation}
where $ \ell^2 =\lambda l_{p}^2$ and $\lambda$ -- dimensionless
numerical factor.
\\GUP, first brought about in a superstring theory \cite{Ven1},
has been supported by the findings in the fields of fundamental
research having no relation to the superstring theory
\cite{GUPg1}--\cite{Kempf}.
\\ Apparently, GUP (\ref{GUP1}) leads to a minimal length
on the order of Planck’s length
\begin{equation}\label{GUP1.1}
\triangle x_{min}=2\sqrt{\lambda}l_{p}
\end{equation}
and, moreover, is responsible for the appearance
of the quantum corrections in the right-hand side of(\ref {GUP4}).
\\ At the present time these corrections are quite clearly understood
(for example, see \cite{mv}--\cite{Nou}), however, leaving some
unanswered questions, the most interesting of which are listed
below.
\\
\\(1) As quantum theory with   GUP  is a deformed quantum theory \cite{Fadd}
with the corresponding deformation  of  the  Heisenberg  algebra
\cite{Magg1},\cite{Kempf}, there is a problem concerning the
dependence of the final result on selection and variability domain
of the deformation parameter.
\\
\\(2) The same problem concerns the entropy of micro(black)holes
or planckian black  holes when we proceed to the region of
planckian  energies.
\\
\\ In this work it is shown that with the use of    $\alpha$,
introduced in \cite{shalyt1}--\cite{shalyt-IJMPD}, as a
deformation parameter a change from lower to higher energies and
vice versa is quite natural, meeting the Hooft-Susskind
Holographic Principle \cite{Hooft1}--\cite{Bou1}, on the one hand,
and the notion of a nonzero minimal energy is also reasonable, on
the other hand.

\section{GUP   and   Quantum Corrections to Semiclassical Formula of BH Entropy}
 As noted in the previous Section, a quantum theory extended by inclusion of GUP is deformed
and associated with a deformation parameter.  Of course, selection
of this parameter is inconclusive (for example,\cite{Magg1}).
\\ In \cite{shalyt-aip}--\cite{shalyt-IJMPD} it has been demonstrated that, at least,
for the simplest variant of GUP – merely deformation of the
commutator $[\vec{x}, \vec{p}]$:
\begin{equation} \label{comm1}
[\vec{x}, \vec{p}]=i\hbar(1+\beta^2\vec{p}^2+...)
\end{equation}
the parameter  $\alpha_{x}=l_{min}^{2}/x^{2}$, where $x$ is the
measuring scale, is also the GUP-deformation parameter, and
(\ref{comm1}) may take the following form (for example,
\cite{shalyt-entropy2}, p. 943):
\begin{equation} \label{comm5}
[\vec{x},\vec{p}]=i\hbar(1+a_{1}\alpha_{x}+a_{2}\alpha_{x}^{2}+...).
\end{equation}
Note that $\alpha_{x}$  may be introduced into a quantum field
theory with the minimal length $l_{min}$, regardless of the origin
of this minimal length (GUP  or
not)\cite{shalyt1}--\cite{shalyt7}.
\\ At the same time, there is one important feature of using the parameter
$\alpha_{x}$  in a theory with the fundamental length (ultraviolet
cutoff). The variability domain of this parameter is
$0<\alpha_{x}\leq 1/4$ \cite{shalyt2}--\cite{shalyt7}and hence,
with this approach based on the high-energy density-matrix
deformation, apart from the {\bf minimal length} $l_{min}$, we
have the {\bf minimal measurable length} $l^{meas}_{min}$
\begin{equation} \label{comm-new1}
l^{meas}_{min}=2l_{min}.
\end{equation}
So, $\alpha_{x}$  is small (for the known energies very small) and
dimensionless. Only at Planck’s scales we have
\begin{equation} \label{comm-new2}
\lim\limits_{x\rightarrow l^{meas}_{min}}\alpha_{x}=\frac{1}{4}.
\end{equation}
The parameter $\alpha_{x}$  is very convenient in studies of the
black hole thermodynamic characteristics and specifically of the
entropy  $S$.
\\ We use the results presented in the work \cite{mv} that is now canonical.
The work gives the most complete formula for the Schwarzshild
black-hole quantum entropy within the scope of GUP that takes the
following form (\cite{mv}, formula (16)):
\begin{equation}\label{GUP5}
S^{BH}_{GUP} =\frac{A}{4l_{p}^{2}}-{\pi\lambda\over 4}\ln
\left(\frac{A}{4l_{p}^{2}}\right) +\sum_{n=1}^{\infty}c_{n}
\left({A\over 4 l_p^2} \right)^{-n}+ \rm{const}\;,
\end{equation}
where the expansion coefficients $c_n\propto \lambda^{n+1}$ can
always be computed to any desired order of accuracy.
\\ Then, for a black hole of the radius $R$, in terms of
$\alpha_{R}$ (\ref{GUP5}) is as follows:
\begin{equation}\label{GUP5.1}
S^{BH}_{GUP}[\alpha_{R}] = 4\pi\lambda\alpha_{R}^{-1}
+{\pi\lambda\over 4}\ln{\alpha_{R}} +\sum_{n=1}^{\infty}
\frac{c_{n}}{(4\pi\lambda)^{n}} \alpha_{R}^{n}+ \rm{const}'\;.
\end{equation}
Here the integration constant is redefined as
$\rm{const}->\rm{const}'+\frac{\pi\lambda}{4}\log\lambda$.
\\
\\ Let us find an extremum of $S^{BH}_{GUP}[\alpha_{R}]$  in the interval
$0<\alpha_{R}\leq1/4$ for different values of the numerical factor
$\lambda$. It is clear that this function has no maximum as for
$\alpha_{R}\rightarrow 0$ (the case of large black holes) the
first (semiclassical) term $\propto \alpha_{R}^{-1}$ becomes
infinitely large to suppress the remaining "tail" in the
right-hand side (\ref{GUP5.1}).  Because of this, of particular
importance is finding of a minimum $S^{BH}_{GUP}[\alpha_{R}]$.
\\
As seen, the minimal element of area
\begin{align}
 (\Delta A)_{min}  \simeq \epsilon l_p^2\Delta p\Delta x\\
\Delta p \simeq \frac{\Delta x}{2\lambda l_p^2}
      \left[1-\sqrt{1-\frac{4\lambda l_p^2}{\Delta x^2}}\right]
\end{align}
is a strictly positive value. Assuming that the entropy of such an
area is equal to $b=\ln2$, for the total entropy we can write the
following equation:
\begin{align}
 \frac{dS}{dA}\simeq \frac{(\Delta S)_{min}}{(\Delta A)_{min}}.
\end{align}
On going from the area to the parameter $\alpha_R=\frac{16\pi\lambda
l_p^2}{A}$  this equation takes a simple form
\begin{align}
\frac{dS}{d\alpha_R}=-\pi\lambda
\frac{2+\sqrt{4-\alpha_R}}{\alpha_R^2}
\end{align}
\\
The right-hand side of the equation is obviously negative for all
$\alpha_R<4$ and for $\lambda>0$. Consequently, the entropy is a
monotonically decreasing function of the parameter $\alpha_R$.
 \begin{figure}[h]
     \centering
         \includegraphics{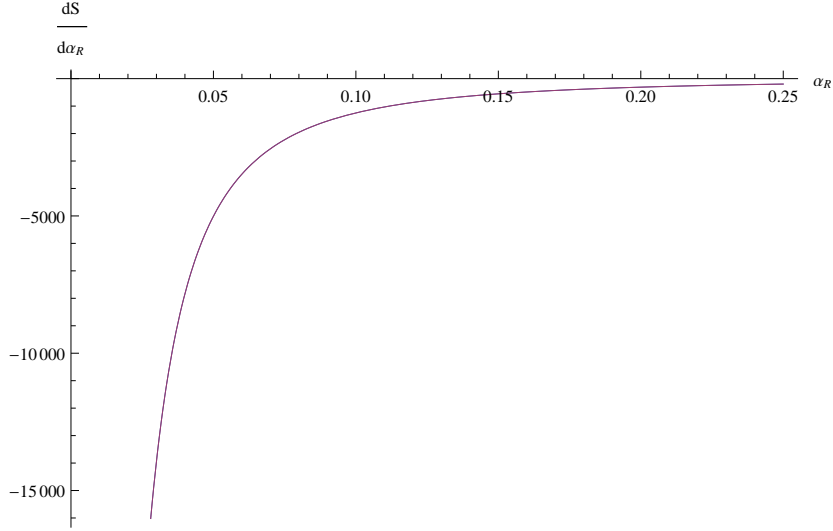}
     \label{fig:dentropy}
\caption{Derivative of the entropy $\frac{dS}{d\alpha_R}$}
 \end{figure}

From the condition $\alpha_R\in (0,1/4]$ it follows that the
entropy is at its minimum at the point $\alpha_R=1/4$.
\\

Integrating the equation, we get
\begin{align}
S &=\pi  \lambda  \left(\frac{4}{\alpha _R}+\frac{\log
\left(\alpha _R\right)}{4} +\frac{\alpha_R}{64}+\frac{\alpha
_R^2}{1024}+\frac{5 \alpha_R^3}{49152}+ \ldots + C
%\frac{29393 \alpha _R^{11}}{387028092977152}+\frac{4199
%   \alpha _R^{10}}{10995116277760}+\frac{2431 \alpha
%   _R^9}{1236950581248}+\frac{715 \alpha _R^8}{68719476736}+\frac{429 \alpha
%   _R^7}{7516192768}+\frac{11 \alpha _R^6}{33554432}+\frac{21 \alpha
%   _R^5}{10485760}+\frac{7 \alpha _R^4}{524288}+
\right).
\end{align}
The integration constant $C$ is fixed in accordance with
\cite{Nozari} from the condition of zero entropy at a minimal area
of the event horizon
$A_p=4\pi(\Delta x_{min})^2=16\pi\lambda l_p^2$.
This complies with the parameter value $\alpha_R=1$. We have derived $C=-4.01672$.
But a minimal \textit{measurable} area is associated with $\alpha_R=1/4$.
This leads to the existence of a minimal entropy that equals $S_{min}=36.5703 \lambda$.
%$S_{min}=49.1892 \lambda$

Performing similar computations for another type of GUP
\begin{align}
 \Delta x\Delta p \geq \sqrt{1+2\lambda l_p^2 \Delta p^2},
\end{align}
we obtain
\begin{align}
 \frac{dS}{d\alpha_R}&=-4\pi\lambda\frac{\sqrt{1-\frac{\alpha_R}{8}}}{\alpha_R^2},
\end{align}

\begin{align}
S &=\pi  \lambda  \left( \frac{4}{\alpha _R}+\frac{\log
\left(\alpha _R\right)}{4} +\frac{\alpha_R}{128}+\frac{\alpha
_R^2}{4096}+\frac{5 \alpha_R^3}{393216}+\ldots + C
%\frac{29393 \alpha
%   _R^{11}}{792633534417207296}+\frac{4199 \alpha
%   _R^{10}}{11258999068426240}+\frac{2431 \alpha
%   _R^9}{633318697598976}+\frac{715 \alpha _R^8}{17592186044416}+\frac{429
%   \alpha _R^7}{962072674304}+\frac{11 \alpha _R^6}{2147483648}+\frac{21
%   \alpha _R^5}{335544320}+\frac{7 \alpha _R^4}{8388608}
\right).
\end{align}
Here the integration constant is $C=-4.008007$, and a minimal entropy
(for $\alpha_R=1/4$) is equal to $S_{min}=36.5911 \lambda$.

It is seen that the difference between these two cases is
insignificant making it possible to infer: the uncertainty
associated with the coefficients $c_{n}$ is  negligible.
 \begin{figure}[h]
     \centering
         \includegraphics{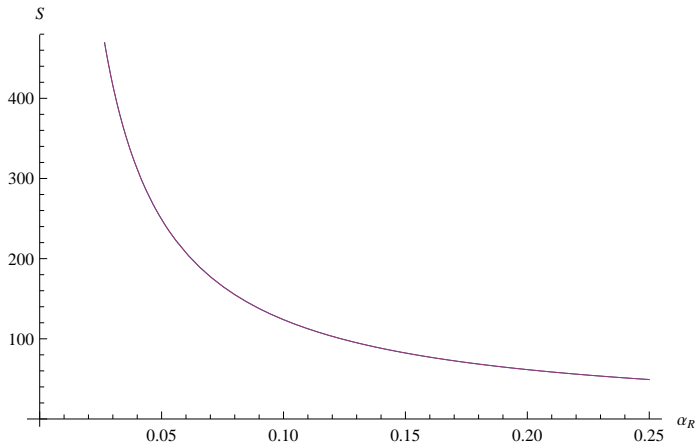}
     \label{fig:entropy}
 \caption{The entropy $S$ as a function of $\alpha_R$}
 \end{figure}

\section{Analysis and Comments}
By our approach, for Schwarzshild black holes there is a {\bf
minimal nonzero entropy}, $S^{bh}_{min}>0$ coming as an attribute
for a micro-black hole of the radius
\begin{equation}\label{bh-1}
r^{min}_{bh}\approx 2l_{min}.
\end{equation}
This result is consistent with the results of other authors, in
particular, in the works \cite{Nicol1}, \cite{Nicol2}, where it
has been demonstrated that the final stage of a black hole is the
planckian remnant with zero entropy and zero temperature.
However, this occurs in the case of the noncommutative space-time,
i.e. when the following condition is fulfilled for operators of
the space coordinates:
\begin{equation}\label{bh-2}
[X^{\mu},X^{\nu}]=i\theta^{\mu\nu},
\end{equation}
where $\theta^{\mu\nu}$ is an anti-symmetric matrix.
\\ This restriction is very strong but it is not imposed in the present paper.
\\ As shown in (\cite{Nozari},section 2), the simplest variant of GUP
(without the noncommutativity condition (\ref{bh-2})) may finally
result in  a zero entropy for the planckian black-hole remnant
(\cite{Nozari}, formulae (21),(22)). However, in \cite{Nozari}, as
distinct from  (\ref{bh-1}), the consideration is given to the
limiting case when
\begin{equation}\label{bh-3}
r^{min}_{bh}=l_{min}=l_{p}.
\end{equation}
The cutoff introduced due to (\ref{comm-new1}),
(\ref{comm-new2}) factors out this case and hence there is no conflict with the result of \cite{Nozari}.
\\
\\ The viewpoint of this work, where the considered “minimal” black holes have a radius that is equal
to the minimal measurable length $l^{meas}_{min}=2l_{min}$
(\ref{comm-new1})) rather than to the minimal length $l_{min}$,
offers particular advantages summarized below.
\\
\\I. This viewpoint enables one to hold for such holes the holographic principle \cite{Hooft1}--\cite{Bou1}(though within the quantum corrections).
It is clear that the zero-entropy planckian remnants violate this
principle cardinally.
\\
\\II. Besides, from the viewpoint of a quantum information theory, such remnants are as good as empty space,
though having great curvature, and this is very odd.
\\
\\III. It is known that micro-black holes are important elements of the presently available models for the space-time foam (for example,
\cite{Scardigli}) but, proceeding from item II, these models are
liable to be problematic.
\\
\\IV. Finally, at least by the heuristic approach, it has been demonstrated that the Holographic Principle
is following from the space-time foam models (for example,
\cite{Ng}), i.e. from Planck’s energies. Because of this, taking
into consideration the inferences of items I--III, it is desirable
to hold this Principle in some or other form at Planck’s scales as
well.

\section{Conclusion}
If evaporation of a black hole results finally in the planckian
remnant with zero energy and temperature, in addition, this means
that Einstein’s equation for the spherically-symmetric horizon
space, close to horizon written as a thermodynamic identity
(first law of thermodynamics) (\cite{Padm13} formula (119)), in
principle has no ultraviolet limit (cutoff) as in this limit the
left-hand side of equation (119) in \cite{Padm13} becomes
degenerate.
\\ As shown in \cite{shalyt-IJMPD}, when such a limit is considered in terms
of   the deformation parameter $\alpha_R$ and its corresponding
variability domain $0<\alpha_R\leq 1/4$ (then called the {\bf
measurable ultraviolet limit}), it is applicable within the scope
of both equilibrium and nonequilibrium thermodynamics.
                               .
%\begin{thebibliography}{000} %for 3 digits
%\begin{thebibliography}{00}  %for 2 digits

\end{document}